\documentclass{ws-mpla}
\usepackage{amsmath}
\newcommand\imag{\mathrm{i}}
\DeclareMathOperator{\tr}{Tr}
\DeclareMathOperator{\re}{Re}

\begin{document}

\markboth{Tom\'{a}\v{s} Brauner}{Two-flavor two-color QCD}
\title{ON THE CHIRAL PERTURBATION THEORY FOR TWO-FLAVOR TWO-COLOR QCD AT
FINITE CHEMICAL POTENTIAL}
\author{\footnotesize TOM\'{A}\v{S} BRAUNER}
\address{Department of Theoretical Physics\\
Nuclear Physics Institute, Academy of Sciences of the Czech Republic\\
\v{R}e\v{z} (Prague), 25068 Czech Republic\\
brauner@ujf.cas.cz}

\maketitle

\pub{Received (Day Month Year)}{Revised (Day Month Year)}
\begin{abstract}
We construct the chiral perturbation theory for two-color QCD with two quark
flavors as an effective theory on the $SO(6)/SO(5)$ coset space. This
formulation turns out to be particularly useful for extracting the physical content
of the theory when finite baryon and isospin chemical potentials are
introduced, and Bose--Einstein condensation sets on.

\keywords{Two-color QCD; chiral perturbation theory; chemical potential.}
\end{abstract}
\ccode{PACS Nos.: 11.30.Rd.}

\section{Introduction}
The phase diagram of quantum chromodynamics (QCD) has attracted much attention
in recent years. The region of high baryon density and low temperature is
relevant for the description of deconfined quark matter, which can be found in
the centers of neutron stars. It is expected to exhibit a~variety of
color-superconducting phases.\cite{Rajagopal:2000wf}

Unfortunately, there is only very little firm knowledge concerning the behavior
of the cold and dense quark matter. At very high densities, asymptotic freedom
of QCD allows one to use weak-coupling methods to determine the structure of
the ground state. On the other hand, the phenomenologically interesting region
of densities corresponds to the strong-coupling regime where \emph{ab initio}
calculations within QCD are not available.

At the same time, current techniques of lattice numerical computations are not
able to reach sufficiently high densities, due to the complexity of the
fermionic Dirac operator that occurs in the Euclidean path-integral measure.
This gave rise to the interest in QCD-like theories that are amenable to
lattice simulations, in particular the two-color QCD with fundamental quarks
and three-color QCD with adjoint quarks.\cite{Kogut:1999iv,Kogut:2000ek}

It turns out that these theories may also be studied by the means of
a~low-energy effective field theory similar to the chiral perturbation theory
of QCD, and non-trivial information about their phase diagram thus obtained.
Within this approach, the structure of the phase diagram has been investigated
in detail, including the effects of finite
temperature.\cite{Kogut:2000ek,Splittorff:2000mm,Splittorff:2001fy,Splittorff:2002xn}
The model-independent predictions of the low-energy effective field theory have
been complemented by lattice computations\cite{Kogut:2001na,Kogut:2003ju} and
calculations within several
models.\cite{Lenaghan:2001sd,Wirstam:2002be,Vanderheyden:2001gx,Ratti:2004ra}

The aim of the present letter is to provide an alternative low-energy effective
formulation of the simplest of this class of theories --- the two-color QCD
with two quark flavors.\footnote{The determinant of the Dirac operator is
always real in two-color QCD, it is, however, positive only for an even number
of flavors.} While the general description of the whole class is based on the
extended $SU(2N_f)$ chiral symmetry of the underlying Lagrangian with $N_f$
quark flavors, we construct the effective Lagrangian by exploiting the Lie
algebra isomorphism $SU(4)\simeq SO(6)$. We show that such a~picture displays
more transparently the physical content of the theory and at the same time
allows for an easy determination of the true ground state, which has been
sought by a~convenient ansatz previously.

The paper is organized as follows. In the next section we summarize the basic
features of two-color QCD to set the stage for the following considerations.
Next we work out the mapping between the coset space $SO(6)/SO(5)$ that we use,
and the $SU(4)/Sp(4)$ used in the literature. The rest of the paper is devoted
to the construction of the effective Lagrangian and its detailed analysis.

\section{Two-color QCD}
In this section we recall the basic properties of two-color QCD, following
closely the treatment of Kogut \emph{et al.}\cite{Kogut:2000ek} The
distinguishing feature of two-color QCD is the pseudoreality of the gauge group
generators, the Pauli matrices, $T_k^*=-T_2T_kT_2$. Consider now a~set of $N_f$
quark flavors in the fundamental representation of the gauge group. As an
immediate consequence, we may trade the right-handed component of the quark
field, $\psi_R$ (flavor and color indices are suppressed), for the left-handed
conjugate spinor $\tilde\psi_R=\sigma_2T_2\psi_R^*$. (The Pauli matrices
$\sigma_k$ act in the Dirac space.)

Instead of the usual Dirac spinor, $\psi=(\psi_L\ \psi_R)^T$, we now work with
the left-handed spinor, $\Psi=(\psi_L\ \tilde\psi_R)^T$, in terms of which the
quark Euclidean Lagrangian of the massive two-color QCD at finite chemical
potential becomes
\begin{equation}
\mathcal L=\imag\Psi^{\dagger}\sigma_{\nu}(D_{\nu}-
\Omega_{\nu})\Psi-m\left[\tfrac12\Psi^T\sigma_2T_2M\Psi+\text{H.c.}\right].
\label{micro_Lagrangian}
\end{equation}

Here $D_{\nu}$ is the gauge-covariant derivative that includes the $SU(2)$
gluon field. $\Omega_{\nu}$ is the static uniform external $U(1)$ gauge field
that incorporates the chemical potential.\cite{Kapusta:1981aa} In the
two-flavor case we shall deal with both the baryon number and the isospin
chemical potential, $\mu_B$ and $\mu_I$, respectively, so that $\Omega_{\nu}$
will eventually be set to $\Omega_{\nu}=\delta_{\nu0}(\mu_BB+\mu_II)$. Here,
\begin{equation}
B=\frac12\left(
\begin{array}{cc}
1 & 0\\
0 & -1
\end{array}\right),\quad
I=\frac12\left(
\begin{array}{cc}
\tau_3 & 0\\
0 & -\tau_3
\end{array}\right)
\label{charge_matrices}
\end{equation}
are the baryon number and isospin generators, respectively. (The Pauli matrices
$\tau_k$ act in the flavor space.) Finally,
$$
M=\left(
\begin{array}{cc}
0 & 1\\
-1 & 0
\end{array}\right)
$$
denotes the mass matrix in the basis of the spinor $\Psi$ and $\sigma_{\nu}$
stands for the four-vector of spin matrices, $\sigma_{\nu}=(-\imag,\sigma_k)$.

In the chiral limit and the absence of the chemical potential, the Lagrangian
Eq. \eqref{micro_Lagrangian} is invariant under the extended global symmetry
$SU(2N_f)$, which includes the naive chiral group $SU(N_f)_L\times SU(N_f)_R$
and additional symmetry transformations due to the pseudoreality of the gauge
group generators. The global symmetry is spontaneously broken by the standard
chiral condensate down to its $Sp(2N_f)$ subgroup.

The low-energy effective field theory for the Goldstone bosons of the broken
symmetry is thus naturally constructed on the coset space $SU(2N_f)/Sp(2N_f)$.
This is parametrized by an antisymmetric unimodular unitary matrix $\Sigma$, in
terms of which the leading-order effective Lagrangian reads
\begin{equation}
\mathcal
L_{\text{eff}}=\frac{F^2}2\tr(\nabla_{\nu}\Sigma\nabla_{\nu}\Sigma^{\dagger})-
G\re\tr(J\Sigma).
\label{eff_Lagrangian}
\end{equation}
The $\nabla$'s denote the covariant derivatives,
\begin{align*}
\nabla_{\nu}\Sigma&=\partial_{\nu}\Sigma-(\Omega_{\nu}\Sigma+\Sigma\Omega^T_{\nu}),\\
\nabla_{\nu}\Sigma^{\dagger}&=\partial_{\nu}\Sigma^{\dagger}+(\Sigma^{\dagger}\Omega_{\nu}+\Omega^T_{\nu}\Sigma^{\dagger}),
\end{align*}
while $J$ serves as a~source field for $\Sigma$, and is eventually set to $mM$.
The quark mass $m$ is connected to the Goldstone boson mass squared $m_{\pi}^2$
by the Gell-Mann--Oakes--Renner relation
$$
mG=F^2m_{\pi}^2.
$$

It is worth emphasizing that the incorporation of the chemical potential into
the effective theory involves no extra free parameters --- the way the chemical
potential enters the Lagrangian is fixed by the form of the covariant
derivatives.

\section{The $SO(6)/SO(5)$ coset space}
From now on we shall restrict our attention to the case $N_f=2$. In that case,
note the Lie algebra isomorphisms $SU(4)\simeq SO(6)$ and $Sp(4)\simeq SO(5)$.
This allows us to recast the low-energy effective field theory on the
$SO(6)/SO(5)$ coset.\cite{Smilga:1995tb} There are altogether five degrees of
freedom, or Goldstone bosons, corresponding to the five independent entries of
the antisymmetric unimodular unitary matrix $\Sigma$.

\subsection{Matrix basis}
The mapping to the $SO(6)/SO(5)$ coset space is now provided by the formula
\begin{equation}
\Sigma=n_i\Sigma_i,
\label{coset_mapping}
\end{equation}
where $\vec n$ is a~six-dimensional real unit vector and $\Sigma_i$ is
a~convenient set of independent antisymmetric $4\times4$ matrices. For $\Sigma$
to be unitary, the basis matrices must satisfy the constraint
\begin{equation}
\Sigma_i^{\dagger}\Sigma_j+\Sigma_j^{\dagger}\Sigma_i=2\delta_{ij}.
\label{OGnality}
\end{equation}
Such a~relation is fulfilled for instance by the matrices
\begin{equation*}
\begin{split}
\Sigma_1=\left(
\begin{array}{cc}
0 & -1\\
1 & 0
\end{array}\right),\quad
\Sigma_2=\left(
\begin{array}{cc}
\tau_2 & 0\\
0 & \tau_2
\end{array}\right),\quad
\Sigma_3=\left(
\begin{array}{cc}
0 & \imag\tau_1\\
-\imag\tau_1 & 0
\end{array}\right),\\
\Sigma_4=\left(
\begin{array}{cc}
\imag\tau_2 & 0\\
0 & -\imag\tau_2
\end{array}\right),\quad
\Sigma_5=\left(
\begin{array}{cc}
0 & \imag\tau_2\\
\imag\tau_2 & 0
\end{array}\right),\quad
\Sigma_6=\left(
\begin{array}{cc}
0 & \imag\tau_3\\
-\imag\tau_3 & 0
\end{array}\right).
\end{split}
\end{equation*}

This particular set has been chosen to comply with existing literature. In
fact, Kogut \emph{et al.}\cite{Kogut:2000ek} use the notation $\Sigma_c$ and
$\Sigma_d$ for our $\Sigma_1$ and $\Sigma_2$, respectively, while Splittorff
\emph{et al.}\cite{Splittorff:2000mm} denote our $\Sigma_1$, $\Sigma_2$ and
$\Sigma_3$ by $\Sigma_M$, $\Sigma_B$ and $\Sigma_I$, respectively.

Let us in addition show a~simple argument that suggests how to choose in
general a~set of matrices satisfying Eq. \eqref{OGnality}. Recall that six
independent antisymmetric \emph{Hermitian} $4\times4$ matrices generate the
real Lie algebra $SO(4)\simeq SO(3)\times SO(3)$. This means that we deal with
two sets of three matrices, which can be shown to fulfill the usual
anticommutator of Pauli matrices, $\{\tau_i,\tau_j\}=2\delta_{ij}$. By
multiplying the matrices from one of the sets by $\imag$, we arrive at three
Hermitian matrices, $H_i=\{\Sigma_2,\Sigma_3,\Sigma_6\}$, and three
anti-Hermitian ones, $A_i=\{\Sigma_1,\Sigma_4,\Sigma_5\}$. These satisfy the
relations
$$
\{H_i,H_j\}=2\delta_{ij},\quad
\{A_i,A_j\}=-2\delta_{ij},\quad
[H_i,A_j]=0,
$$
that are equivalent to Eq. \eqref{OGnality}.

\subsection{Structure of the coset}
It remains to prove that Eq. \eqref{coset_mapping} provides a~one-to-one
parametrization of the coset $SU(4)/Sp(4)$. To that end, note that any
antisymmetric $4\times4$ matrix $U$ may be expanded in the basis $\Sigma_i$,
$U=z_i\Sigma_i$, where $z_i$ are in general complex coefficients.
The unitarity of $U$ constrains these coefficients as
$$
1=U^{\dagger}U=\sum_i|z_i|^2+\imag\sum_{i\neq
j}(x_iy_j-x_jy_i)\Sigma^{\dagger}_i\Sigma_j,
$$
the $x_i$ and $y_i$ being the real and imaginary parts of $z_i$, respectively.

It is now crucial to observe that the products $\imag\Sigma^{\dagger}_i\Sigma_j$ for
$i\neq j$ span the set of $15$ linearly independent generators of $SU(4)$ so
that the unitarity of $U$ requires separately $\sum_i|z_i|^2=1$ and
$x_iy_j=x_jy_i$ for all pairs of $i,j$.

The latter condition means that the complex phases of all the $z_i$'s must be
equal so that $z_i=n_ie^{\imag\varphi}$ with real $n_i$, while the former one
requires $\sum_in_i^2=1$. It is a~matter of simple algebra to calculate the
determinant of $U$,
$$
\det U=e^{4\imag\varphi}\biggl(\sum_in_i^2\biggr)^2=e^{4\imag\varphi}.
$$

Since the elements of the coset $SU(4)/Sp(4)$ are unimodular matrices, we are
left with two distinct possibilities, $\varphi=0$ or $\varphi=\pi/2$. (The next
solution, $\varphi=\pi$, already corresponds to $\varphi=0$ with just the sign
of all the $n_i$'s inverted.)

In conclusion, every antisymmetric unimodular unitary matrix $\Sigma$ may be
cast in the form \eqref{coset_mapping}, where $\vec n$ is either real, or pure
imaginary vector. However, in the standard coset construction of the effective
Lagrangian,\cite{Coleman:1969sm,Callan:1969sn} the global symmetry group is
required to act transitively on the parameter space of the Goldstone fields
that is, the actual coset space must be connected. As the chiral condensate,
above which we build our effective theory, is described by the matrix
$\Sigma_1$ (note that $M=-\Sigma_1$), we have to choose the connected component
with real $\vec n$, as in Eq. \eqref{coset_mapping}.

\subsection{Physical content of the basis matrices}
It is instructive to look at the transformation properties of the matrix
$\Sigma$. This will allow us to classify the Goldstone modes by their baryon
and isospin quantum numbers.

Recall that $\Sigma$ is an antisymmetric tensor under $SU(4)$ that is, it
transforms as $\Sigma\to U\Sigma U^T$ for $U\in SU(4)$. For an infinitesimal
transformation generated by the baryon number or the third component of the
isospin we get
$$
\delta_{\varepsilon}\Sigma=\imag\varepsilon(Q\Sigma+\Sigma Q^T)=
\imag\varepsilon\{Q,\Sigma\},\quad Q=B,I.
$$

Let $\Sigma$ be a~general block matrix of the form $\left(\begin{smallmatrix}
K~& L \\ M & N \end{smallmatrix}\right)$. Then
$$
\{B,\Sigma\}=\left(
\begin{array}{cc}
K~& 0\\
0 & -N
\end{array}\right),\quad
\{I,\Sigma\}=\left(
\begin{array}{cc}
\tfrac12\{\tau_3,K\} & \tfrac12[\tau_3,L]\\
-\tfrac12[\tau_3,M] & -\tfrac12\{\tau_3,N\}
\end{array}\right).
$$
The quantum numbers of the particular components of $\Sigma$ are summarized in
Fig. \ref{tab:condensates}.
\begin{figure}
\begin{center}
\begin{tabular}{|c|c|c|}
\hline
$\Sigma_2,\Sigma_4$ & $B=\pm1;I=0$ & diquark and antidiquark\\
\hline
$\Sigma_3,\Sigma_5,\Sigma_6$ & $B=0;I=\pm1,0$ & isospin triplet $\vec\pi$\\
\hline
$\Sigma_1$ & $B=0;I=0$ & singlet $\sigma$\\
\hline
\end{tabular}
\end{center}
\caption{Quantum numbers of the components of the matrix $\Sigma$ in the
expansion \eqref{coset_mapping}.}
\label{tab:condensates}
\end{figure}

To gain more insight into the nature of the effective field $\Sigma$, let us
assign to it a~composite field,
$$
\Sigma\to\tfrac12\Psi^T\sigma_2T_2\Sigma\Psi+\text{H.c.},
$$
which corresponds to the form of the mass term in Eq. \eqref{micro_Lagrangian}.
Such a composite operator may be regarded as an interpolating field for the
Goldstone boson.

With the explicit knowledge of the matrices $\Sigma_i$ it is now
straightforward to find the particle content of the corresponding interpolating
fields, cf. also Fig. \ref{tab:condensates},
\begin{gather*}
\Sigma_2\to-\tfrac12\psi^TC\gamma_5T_2\tau_2\psi+\text{H.c.},\quad
\Sigma_4\to-\tfrac12\imag\psi^TC\gamma_5T_2\tau_2\psi+\text{H.c.},\\
\Sigma_3\to-\imag\bar\psi\tau_1\gamma_5\psi,\quad
\Sigma_5\to\imag\bar\psi\tau_2\gamma_5\psi,\quad
\Sigma_6\to-\imag\bar\psi\tau_3\gamma_5\psi,\\
\Sigma_1\to\bar\psi\psi.
\end{gather*}

\section{Chiral perturbation theory}
We are now ready to write down the leading-order effective Lagrangian and use
it to analyze the phase diagram of the theory. First, we have to minimize the
static part of the Lagrangian in order to determine the ground state at nonzero
chemical potential.

\subsection{Global minimum of the static Lagrangian}
From Eq. \eqref{eff_Lagrangian} we can immediately infer the static part,
\begin{equation}
\mathcal
L_{\text{stat}}=-\frac{F^2}2\tr\left[(\Omega_{\nu}\Sigma+\Sigma\Omega_{\nu}^T)
(\Sigma^{\dagger}\Omega_{\nu}+\Omega^T_{\nu}\Sigma^{\dagger})\right]-G\re\tr(J\Sigma).
\label{static_Lagrangian}
\end{equation}
We include the external source $J$ in the general form
$$
J=j_i\Sigma^{\dagger}_i,
$$
with real $j_i$. Note that setting $j_1=m$, we reproduce the quark mass
contribution to the effective Lagrangian.

The other sources can be taken as infinitesimally small, since they essentially
serve to generate the ground-state condensates,
$$
\langle\Sigma_i\rangle=-\frac{\partial\mathcal L_{\text{stat}}}{\partial j_i}.
$$
From the orthogonality property,
$\tr(\Sigma_i^{\dagger}\Sigma_j)=4\delta_{ij}$,
we find $\re\tr(J\Sigma)=4\vec j\cdot\vec n$ so that we have
$$
\langle\vec\Sigma\rangle=4G\vec n.
$$

It is obvious that the vacuum condensate rotates on a~sphere in the
six-dimensional space, with coordinates corresponding to the six basis matrices
$\Sigma_i$. It remains to calculate the vector $\vec n$ minimizing the static
Lagrangian \eqref{static_Lagrangian}.

Note first that in the absence of chemical potential, $\Omega_{\nu}=0$, the
static Lagrangian is minimal when the condensate is aligned with the external
source $\vec j$.

When baryon and isospin chemical potentials are switched on, we shall for
simplicity assume that only the sources $j_1,j_2,j_3$ are present. This is
sufficient to include the quark mass effects and calculate both the diquark and
the isospin (pion) condensate. Taking into account the explicit form of the
charge matrices, Eq. \eqref{charge_matrices}, the static Lagrangian becomes
$$
\mathcal
L_{\text{stat}}=-2F^2\left[\mu_B^2(n_2^2+n_4^2)+\mu_I^2(n_3^2+n_5^2)\right]-4G(j_1n_1+j_2n_2+j_3n_3).
$$

The first term is invariant under $SO(2)\times SO(2)$ rotations in the planes
$(2,4)$ and $(3,5)$. In the absence of the external sources, this symmetry may
be exploited to set $n_4=n_5=0$. The source $J$ breaks the symmetry and, in
fact, prefers the solutions with $n_4=n_5=0$. The problem of finding the ground
state thus reduces to minimizing the expression,
\begin{equation}
\mathcal
L_{\text{stat}}=-2F^2(\mu_B^2n_2^2+\mu_I^2n_3^2)-4G(j_1n_1+j_2n_2+j_3n_3),
\label{reduce_static_Lagrangian}
\end{equation}
on the sphere $S^5:\vec n^2=1$.

The Lagrangian now does not depend on $n_4,n_5,n_6$ so that we are actually
looking for a~minimum on the ball, $n_1^2+n_2^2+n_3^2\leq1$. It is clear that
at the \emph{global} minimum, both terms on the right hand side of Eq.
\eqref{reduce_static_Lagrangian} are negative (otherwise we could lower the
energy by the inversion, $\vec n\to-\vec n$). By the same token, the global
minimum must lie on the surface of the ball, since if this were not the case,
we could lower the energy by scaling up the vector: $\vec n\to t\vec
n,t>1$.

We have thus shown that in the global minimum, $n_1^2+n_2^2+n_3^2=1$ and
$n_4=n_5=n_6=0$, and the ground-state condensate is given by the linear
combination $\Sigma=n_1\Sigma_1+n_2\Sigma_2+n_3\Sigma_3$. We stress the
simplicity of the proof of this fact within the $SO(6)/SO(5)$ coset formulation
of the chiral perturbation theory. Indeed, using the standard $SU(4)/Sp(4)$
formalism, Splittorff \emph{et al.}\cite{Splittorff:2000mm} only assumed such
a~form of $\Sigma$, and also did not prove that the minimum thus found was global.

To demonstrate the power of the formalism we have built so far, we shall next
rederive the results of Kogut \emph{et al.}\cite{Kogut:2000ek} for the case of
nonzero baryon chemical potential $\mu_B$. We shall thus set $\mu_I=0$ and
$j_i=\delta_{i1}m$. Isospin chemical potential can be introduced along the same
lines and the results of Splittorff \emph{et al.}\cite{Splittorff:2000mm} would
be easily recovered.

With the assumptions made, the static Lagrangian becomes
$$
\mathcal L_{\text{stat}}=-2F^2m_{\pi}^2(x^2\sin^2\alpha+2\cos\alpha),
$$
where $x=\mu_B/m_{\pi}$ and $\alpha$ parametrizes the minimum,
$\Sigma=\Sigma_1\cos\alpha+\Sigma_2\sin\alpha$. (The same argument as above
tells us that when $\mu_I=0$ and $j_3=0$, then $n_3=0$ in the global minimum.)

Now when $x<1$, the minimum occurs at $\alpha=0$ --- only the chiral condensate
is nonzero, this is the normal phase. When, on the other hand, $x>1$, the
Lagrangian is minimized by $\cos\alpha=1/x^2$. In this case, the chiral
condensate rotates into the diquark condensate as the angle $\alpha$ increases.
This is the Bose--Einstein condensation phase.

\subsection{Excitation spectrum}
The spectrum of excitations above the ground state is determined by the
bilinear part of the Lagrangian. Expanding Eq. \eqref{eff_Lagrangian} in terms
of the components $n_i$, it acquires the form
\begin{equation}
\mathcal L_{\text{eff}}=2F^2(\partial_{\nu}\vec
n)^2+4\imag F^2\mu_B(n_2\partial_0n_4-n_4\partial_0n_2)-2F^2\mu_B^2(n_2^2+n_4^2)-4F^2m_{\pi}^2n_1.
\label{bilinear_Lagrangian}
\end{equation}
To proceed, we have to deal separately with the two phases of the theory.

\subsubsection{The normal phase}
When $x<1$, the ground state expectation values of $\vec n$ are $n_1=1$ and all
other components zero. The independent excitations above the ground state may
be identified with $n_i,i=2,\dotsc,6$, while $n_1$ is expressed in terms of
them via the constraint $\vec n^2=1$,
$$
n_1=\sqrt{1-\sum_{i=2}^6n_i^2}=1-\tfrac12\sum_{i=2}^6n_i^2+\text{higher order
terms}.
$$

The bilinear part of the Lagrangian \eqref{bilinear_Lagrangian} becomes
$$
\frac{\mathcal L_{\text{bilin}}}{2F^2}=\sum_{i=3,5,6}(\partial_{\nu}
n_i)^2+(\partial_0 N-\mu_BN)(\partial_0 N^{\dagger}+\mu_BN^{\dagger})+\nabla
N\cdot\nabla N^{\dagger}+ m_{\pi}^2\sum_{i=2}^6n_i^2,
$$
where we have introduced $N=n_2+\imag n_4$, a~complex field that carries baryon
number one. This field thus corresponds to the diquark, while $N^{\dagger}$
describes the antidiquark.

We find the following dispersion relations,
\begin{align*}
E&=\sqrt{\vec p^2+m_{\pi}^2}&&\text{pion triplet $n_3,n_5,n_6$},\\
E&=\sqrt{\vec p^2+m_{\pi}^2}-\mu_B&&\text{diquark $N$},\\
E&=\sqrt{\vec p^2+m_{\pi}^2}+\mu_B&&\text{antidiquark $N^{\dagger}$}.
\end{align*}

\subsubsection{The Bose--Einstein condensation phase}
For $x>1$, the chiral condensate alone is no longer the proper ground state and
the Bose--Einstein condensation sets. We therefore parametrize the field $\vec
n$ as
$$
\vec n=(\rho\cos\varphi,\rho\sin\varphi,n_3,n_4,n_5,n_6).
$$

The ground state corresponds to $\rho=1$, $\varphi=\alpha$ and
$n_3=n_4=n_5=n_6=0$. We set $\varphi=\alpha+\theta$ so that the five independent
degrees of freedom are now $\theta$ and $n_i,i=3,\dotsc,6$. The radial
parameter $\rho$ is given by
$$
\rho=\sqrt{1-\sum_{i=3}^6n_i^2}=1-\tfrac12\sum_{i=3}^6n_i^2+\text{higher order
terms}.
$$

The bilinear Lagrangian reads in this case,
$$
\frac{\mathcal
L_{\text{bilin}}}{2F^2}=\sum_{i=3}^6(\partial_{\nu}n_i)^2+(\partial_{\nu}\theta)^2+
2\imag\mu_B(\theta\partial_0n_4-n_4\partial_0\theta)\cos\alpha+\mu_B^2\biggl(\sum_{i=3,5,6}n_i^2+\theta^2\sin^2\alpha\biggr).
$$
Three of the degrees of freedom, $n_3,n_5,n_6$, again represent the pion
triplet, now with the dispersion relation $E=\sqrt{\vec p^2+\mu_B^2}$. The
dispersions of the remaining two excitations are obtained by a diagonalization
of the inverse propagator in the $(\theta,n_4)$ sector. The result is
$$
E_{\pm}^2=\vec p^2+\frac{\mu_B^2}2(1+3\cos^2\alpha)\pm\frac{\mu_B}2\sqrt{\mu_B^2(1+3\cos^2\alpha)^2+16\vec
p^2\cos^2\alpha},
$$
in accord with previous work.\cite{Kogut:2000ek,Splittorff:2000mm} The masses
of these modes are given by
$$
m_+^2=\mu_B^2(1+3\cos^2\alpha)=\mu_B^2+\frac{3m_{\pi}^2}{\mu_B^2},\quad
m_-^2=0.
$$

In contrast to the normal phase, there is always one truly massless Goldstone
boson stemming from the exact baryon number $U(1)$ symmetry, which is
spontaneously broken by the diquark condensate $\Sigma_2$.

It is, however, worth emphasizing that the nature of this Goldstone boson, as
well as of the massive mode, changes as the chemical potential increases. There
are two reasons --- the rotation of the ground state in the $(n_1,n_2)$ plane, and the
balance between the mass term in the bilinear Lagrangian and the term with
a~single time derivative.

In the limit $\alpha\to0$, the parameter $\theta$ is, to the lowest order,
equal to $n_2$ and the parametrization of the case $x<1$ is recovered. Here the
Goldstone boson is the diquark $N=\theta+in_4$.

\begin{figure}
\begin{center}
\includegraphics{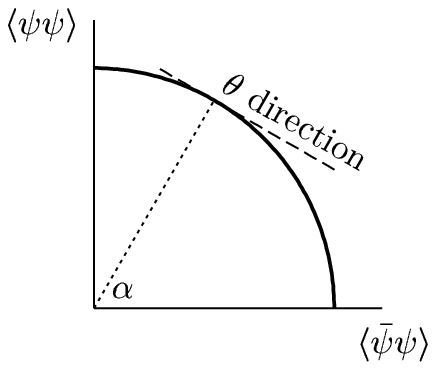}
\end{center}
\caption{The orientation of $\theta$ is perpendicular to the direction of the
ground-state condensate, which is represented by the dotted line. The
coordinates $n_1,n_2$ are labeled schematically by the chiral and the diquark
condensate, respectively.}
\label{fig:circle}
\end{figure}
As the angle $\alpha$ grows, the orientation of $\theta$ also changes in the
$(n_1,n_2)$ plane so that it is always perpendicular to the direction of the
condensate, see Fig. \ref{fig:circle}. In the limit $\alpha\to\pi/2$ that is,
$\mu_B\gg m_{\pi}$, the condensate is purely diquark and $\theta$ has the
quantum numbers of $n_1$, i.e. the $\sigma$ field. The Goldstone boson is now
$n_4$. Note also that it is a~linear combination of the diquark and the
antidiquark so that it has no definite baryon number. This is, of course, not
surprising since the baryon number is spontaneously broken and thus it cannot
be used to label the physical states.

\section{Conclusions}
We have constructed the chiral perturbation theory for two-color QCD with two
quark flavors on the $SO(6)/SO(5)$ coset. We have provided an explicit mapping
between this formulation and that used previously in literature, based on the
$SU(4)/Sp(4)$ coset space.

The virtue of the present approach is that the orthogonal rotations, in
contrast to the unitary symplectic transformations, can be easily visualized
and the physical content of the theory thus made manifest. We were also able to
give a~simple proof of the fact that the condensate taken previously as an
ansatz is indeed the true ground state, and we thus justified the assumptions
made in the older work.

Since the $SO(N)$-symmetric nonlinear sigma model is known to great detail, the
connection provided in this letter can hopefully lead to the improvement in the
understanding of the phase diagram of the two-color QCD at low energies, at
least in the two-flavor case.

\section*{Acknowledgments}
The author is grateful to Ji\v{r}\'{\i} Ho\v{s}ek for critical reading of the
manuscript. The present work was supported in part by the Institutional
Research Plan AV0Z10480505, and by the GACR grant No. 202/05/H003.


\begin{thebibliography}{0}

\bibitem{Rajagopal:2000wf}
K.~Rajagopal and F.~Wilczek, in {\it At the Frontier of Particle Physics: Handbook of QCD},
ed.~M.~Shifman (World Scientific, 2001).

\bibitem{Kogut:1999iv}
J.~B. Kogut, M.~A. Stephanov, and D.~Toublan, {\it Phys. Lett.} {\bf B464}, 183--191 (1999).

\bibitem{Kogut:2000ek}
J.~B. Kogut, M.~A. Stephanov, D.~Toublan, J.~J.~M. Verbaarschot, and
  A.~Zhitnitsky, {\it Nucl. Phys.} {\bf B582}, 477--513 (2000).

\bibitem{Splittorff:2000mm}
K.~Splittorff, D.~T. Son, and M.~A. Stephanov, {\it Phys. Rev.} {\bf D64}, 016003 (2001).

\bibitem{Splittorff:2001fy}
K.~Splittorff, D.~Toublan, and J.~J.~M. Verbaarschot, {\it Nucl. Phys.} {\bf
  B620}, 290--314 (2002).

\bibitem{Splittorff:2002xn}
K.~Splittorff, D.~Toublan, and J.~J.~M. Verbaarschot, {\it Nucl. Phys.} {\bf B639}, 524--548 (2002).

\bibitem{Kogut:2001na}
J.~B. Kogut, D.~K. Sinclair, S.~J. Hands, and S.~E. Morrison, {\it Phys. Rev.} {\bf
D64}, 094505 (2001).

\bibitem{Kogut:2003ju}
J.~B. Kogut, D.~Toublan, and D.~K. Sinclair, {\it Phys. Rev.} {\bf D68}, 054507 (2003).

\bibitem{Lenaghan:2001sd}
J.~T. Lenaghan, F.~Sannino, and K.~Splittorff, {\it Phys. Rev.} {\bf D65}, 054002 (2002).

\bibitem{Wirstam:2002be}
J.~Wirstam, J.~T. Lenaghan, and K.~Splittorff, {\it Phys. Rev.} {\bf D67}, 034021 (2003).

\bibitem{Vanderheyden:2001gx}
B.~Vanderheyden and A.~D. Jackson, {\it Phys. Rev.} {\bf D64}, 074016 (2001).

\bibitem{Ratti:2004ra}
C.~Ratti and W.~Weise, {\it Phys. Rev.} {\bf D70}, 054013 (2004).

\bibitem{Kapusta:1981aa}
J.~I. Kapusta, {\it Phys. Rev.} {\bf D24}, 426--439 (1981).

\bibitem{Smilga:1995tb}
A.~Smilga and J.~J.~M. Verbaarschot, {\it Phys. Rev.} {\bf D51}, 829--837 (1995).

\bibitem{Coleman:1969sm}
S.~R. Coleman, J.~Wess, and B.~Zumino, {\it Phys. Rev.} {\bf 177}, 2239--2247 (1969).

\bibitem{Callan:1969sn}
C.~G. Callan, S.~R. Coleman, J.~Wess, and B.~Zumino, {\it Phys. Rev.} {\bf
177}, 2247--2250 (1969).

\end{thebibliography}
\end{document}